\documentclass[preprint,showpacs,preprintnumbers,amsmath,amssymb]{revtex4}

\usepackage{graphicx}
\usepackage{dcolumn}
\usepackage{bm}

\newcommand{\be}{\begin{equation}}
\newcommand{\en}{\end{equation}}
\newcommand{\bea}{\begin{eqnarray}}
\newcommand{\ena}{\end{eqnarray}}

\begin{document}


\title{Emergent universe scenario and the low CMB multipoles}

\author{Pedro Labra\~{n}a}
 \email{plabrana@ubiobio.cl}
\affiliation{Departamento de F\'{i}sica, Universidad del B\'{i}o-B\'{i}o, Casilla 5-C, Concepci\'on, Chile and \\
Departament d'Estructura i Constituents de la Mat\`{e}ria, Institut
de Ci\`{e}ncies del Cosmos, Universitat de Barcelona, Diagonal 647,
08028 Barcelona, Spain.}

\date{\today}

\begin{abstract}

In this work we study superinflation in the context of the emergent
universe (EU) scenario.
The existence of a superinflating phase before the onset of
slow-roll inflation arises in any emergent universe model.
We found that the superinflationary period in the EU scenario
produces a suppression of the CMB anisotropies at large scale which
could be responsible for the observed lack of power at large angular
scales of the CMB.

\end{abstract}

\maketitle

\section{Introduction}
\label{Int}

Cosmological inflation has become an integral part of the standard
model of the Universe. Apart from being capable of removing the
shortcomings of the standard cosmology, it gives important clues for
large scale structure formation \cite{Guth1, Albrecht, Linde1,
Linde2} (see \cite{libro} for a review).

The scheme of inflation is based on the idea that there was an early
phase, before the big bang, in which the Universe evolved through a
nearly exponential expansion during a short period of time at
high-energy scales. During this phase, the Universe was dominated by
a potential $V(\phi)$ of a scalar field, which is called the
inflaton $\phi$.
The idea of cosmological inflation was first developed by
Guth~\cite{Guth1} and later refined simultaneously by
Linde~\cite{Linde1, Linde2} and Albrecht and
Steinhardt~\cite{Albrecht} in a version known as slow-roll
inflation.

In the standard inflationary universe quantum fluctuations of the
inflaton field give rise to a curvature perturbation that is
constant for modes outside the horizon. This curvature perturbation
is then the seed for structure formation in the Universe. The
quantum fluctuations of the inflaton could be calculated using the
semiclassical theory of quantum fields in curved spacetime; see
\cite{Mukhanov, Hawkin,Guth, Staro, Bardeen}.
In particular, for a very flat inflaton potential, the inflaton can
be taken to be massless and the quantum fluctuation becomes
$\delta\phi(k) = H_k/(2\pi)$ at the moment of crossing the horizon,
where $H$ is the Hubble parameter during inflation and $k$ is the
comoving momentum related with the fluctuation.
The notation on the right-hand side emphasizes the following
property: Modes of different momenta exit the horizon at different
times, with the larger $k$ being the later.
Once a given mode exits the horizon, its amplitude freeze. This
perturbation then reenters the horizon during the postinflationary
era and becomes the seed for the  structure formation, leaving its
imprint on the CMB.

The standard slow-roll inflation predicts a slightly red-tilted
power spectrum of the primordial perturbation. This means that large
scale (small $k$) perturbation has more power than small scale
perturbation.
This red tilt arises from the fact that the Hubble parameter slowly
decreases with time as the scalar field rolls downwards to its
potential.

The recent cosmological observations are entirely consistent with
the simplest slow-roll inflationary models \cite{WMAP1, planckI,
PlanckXXII}. Probing that the observed universe is almost flat, the
bispectral non-Gaussianity parameter $f_{NL}$ measured  is
consistent with zero. The scalar perturbation spectral index $n_s$
is less than one, which is a measure of the tilt discussed above
($n_s <1$ red tilt, $n_s >1$ blue tilt and $n_s =1$ scale invariant
spectrum). All these results are predicted by the simplest slow-roll
inflationary models.

However, there are intriguing observations on cosmic microwave
background radiation, suggesting a lack of power at large angular
scales (very low multipoles, $l < 40$). These results were first
obtained by COBE \cite{cobe} and WMAP \cite{WMAP1} and now are
confirmed by Planck \cite{planckI}.
Although these results are well within our cosmic variance and
statistically their significance is still low, the power deficit is
not insignificant.
The Planck Collaboration reported a power deficit in the low
multipoles CMB power spectrum of order $5\% - 10 \%$ (with respect
to the Planck best-fit $\Lambda$CDM model \cite{PlanckXV,
PlanckXXIII}) with statistical significance $2.5 \sim 3 \,\sigma$.

This situation is interesting because the very low $l$ modes in the
CMB spectrum at present time, correspond to very large wavelength
modes. Since these modes have been superhorizon sized between
inflation and now, they have not been contaminated by the later
evolution of the Universe. For this reason, we could attribute the
new feature observed in the spectrum at low $l$ to physics at the
very earliest Universe, perhaps before slow-roll inflation
\cite{BasteroGil:2003bv}.

There are different approaches developed in order to explain this
problem. One possibility is to consider the hypothesis of the
\emph{small universe} with a compact topology.
In this case, perturbations on scales exceeding the fundamental cell
size are suppressed; see \cite{top1, top2, top3, top4, top5, top6,
top7}.
Another possibility is hyperspherical topology corresponding to a
closed universe \cite{top8, top9}, or consider an anisotropic
universe \cite{Campanelli:2006vb,Campanelli:2007qn,
Campanelli:2009tk}.
The topological approach has some problems with the S-statistic and
the matched circle test \cite{Zaldarriaga}. On the other hand,
Planck searches yield no detection of the compact topology
\cite{PlanckXXVI}.

Another approach is to introduce a cutoff in the primordial power
spectrum \cite{cut1, cut2, Contaldi:2003zv,cut4, cut5, cut6,
Ramirez:2011kk, Dudas:2012vv, Piao:2003zm, Liu:2013kea}. This cutoff
is normally introduced by hand but linked to the spatial curvature
scale \cite{top9}, string physics \cite{cut6, Dudas:2012vv,
Piao:2003zm}, the bouncing universe scheme \cite{Piao:2003zm,
Liu:2013kea}, or a fast-rolling stage in the evolution of the
inflaton field \cite{Contaldi:2003zv}.
This approach is interesting since it has been claimed that from the
observed angular power spectrum, it is possible to deconvolve the
primordial power spectrum for a given set of cosmological
parameters.
The most prominent feature of the recovered primordial power
spectrum is a sharp, infrared cutoff on the horizon scale
\cite{cutwmap1, cutwmap2}.

In this respect, it has been suggested that the low-$l$ power could
be related to a period of superinflation, previous to the standard
slow-roll inflationary regimen \cite{Zhang:2003eh, Biswas:2013dry,
Liu:2013iha}, where the superinflationary period is characterized by
the condition $\dot{H}
>0$.
In particular, in Ref.~\cite{Biswas:2013dry}, this possibility was
studied in the context of bouncing universes and was discussed
regarding the emergent universe scheme.

It is interesting to note, see \cite{Biswas:2013dry}, that a
superinflationary period is related to any mechanism which attempts
to solve the cosmological singularity problem \cite{Borde:1993xh,
Borde:1997pp, Guth:1999rh, Borde:2001nh, Vilenkin:2002ev} in a
semiclassical spacetime description.
There are two ways to avoid the singularity problem in this context.
One possibility is to consider a nonsingular bounce
\cite{Cai:2007qw, Biswas:2005qr, Peter:2002cn, Novello:2008ra,
BT1,BT2,BT3,BT4,BT5, BT6,Singh:2006im,
Battefeld:2004mn,Gasperini:2003pb,Hwang:2001zt,
bounce0,bounce1,bounce2,bounce3, bounce4, Liu:2013kea};
the other possibility is to consider the emergent universe scenario
\cite{Ellis:2002we,Ellis:2003qz, Mulryne:2005ef}.


The emergent universe (EU) refers to models in which the universe
emerges from an Einstein static state, inflates, and is then
submitted into a hot big bang era. Such models are appealing since
they provide specific examples of non–singular (geodesically
complete) inflationary universes. These models have been studied in
different contexts during recent years; see \cite{Ellis:2002we,
Ellis:2003qz, Mulryne:2005ef, Mukherjee:2005zt, Mukherjee:2006ds,
Banerjee:2007qi, Nunes:2005ra, Lidsey:2006md, Beesham:2009zw,
Paul:2008id, Debnath:2008nu, Banerjee:2007sg, delCampo:2007mp,
delCampo:2009kp, delCampo:2010kf,delCampo:2011mq,
Labrana:2011np,Cai:2012yf, Rudra:2012mu, Liu:2012ww,
Guendelman:2013dka, Cai:2013rna}.

In this paper, we study the period of superinflation in the context
of the EU scenario.
The existence of a superinflating phase before the onset of
slow-roll inflation arises in any EU scenario \cite{Biswas:2013dry}
and has not been studied thus far.
As an arena to explore the superinflation phase, we consider the EU
model developed in Ref.~\cite{Ellis:2002we, Ellis:2003qz}. This EU
model is based on general relativity and considered a closed
universe dominated by a scalar field minimally coupled to gravity.
We found that the superinflationary period in the EU scenario
produces a suppression of the CMB anisotropies at large scale which
could be responsible of the observed lack of power at large angular
scales of the CMB \cite{cobe, WMAP1, planckI}.


The paper is organized as follows. In Sec.~\ref{EU} we present the
principal characteristics of the EU scenario and the
superinflationary regimen.
In Sec.~\ref{Perturbation}  we calculate the primordial perturbation
generated during the superinflationary phase in the context of the
EU scenario.
In Sec.~\ref{Effective-Power} we estimate the effective power
spectrum generated by the EU model which takes into account the
early period of superinflation and the subsequent period of standard
slow roll.
In Sec.~\ref{CMB-Anisotropies} we compute the theoretical CMB power
spectrum of this model by using CLASS code \cite{Lesgourgues:2011re,
Blas:2011rf}.
In Sec.~\ref{Sec-conclusions} we summarize our results.

\section{The Emergent Universe scenario}\label{EU}

In the emergent universe scenario, the Universe is initially in a
past-eternal classical Einstein static state which eventually
evolves into a subsequent inflationary phase; see
\cite{Ellis:2002we,Ellis:2003qz,Mulryne:2005ef,Mukherjee:2005zt,
Mukherjee:2006ds,Banerjee:2007qi,Nunes:2005ra,Lidsey:2006md}.
During the past-eternal static regime, it is assumed that the scalar
field is rolling on the asymptotically flat part of the scalar
potential with a constant velocity, providing the conditions for a
static universe. But once the scalar field exceeds some value, the
scalar potential slowly drops from its original value. The overall
effect of this is to distort the equilibrium behavior breaking the
static solution.
If the potential has a suitable form in this region, slow-roll
inflation will occur.


\begin{figure}
\centering
\includegraphics[width=10cm]{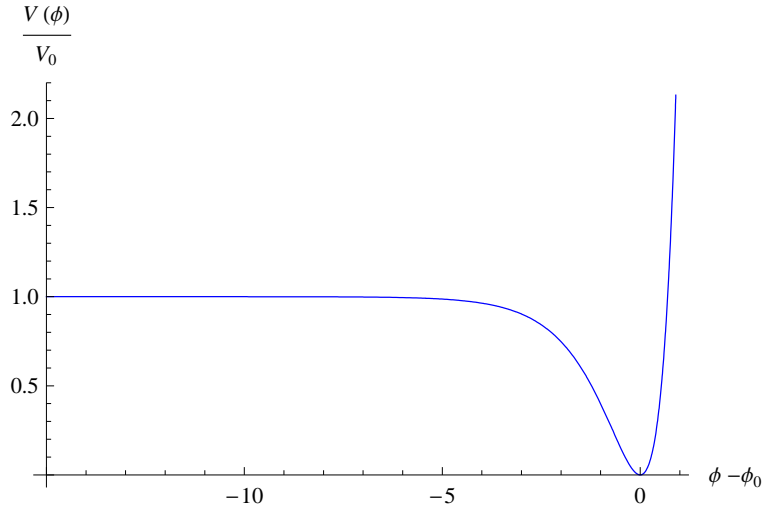}
\caption{Schematic representation of a potential for a standard
emergent universe scenario. \label{Potencial-1}}
\end{figure}

In the EU scenario, the evolution of the scale factor could be
modeled by the following expression (see\cite{Ellis:2002we,
Ellis:2003qz}):
\begin{equation}\label{a-EU1}
a(t) \simeq a_0 + A\,e^{H_0\,t}
\end{equation}
where $a_0, A, H_0$ are positive constants. This universe is past
asymptotic to an Einstein static state, since $a(t)\rightarrow a_0$
as $t \rightarrow -\infty$. Thus, $a_0$ is identified with the
radius of the Einstein static universe. At late times, on the other
hand, $a(t)\rightarrow  A\,e^{H_0\,t}$ and the model approach a de
Sitter expansion phase.

For example, in Ref.~\cite{Ellis:2003qz}, the scalar potential has
been reconstructed from the evolution Eq.~(\ref{a-EU1}), where we
can note that this reconstructed potential exhibits the same shape
of the effective potential Fig.~\ref{Potencial-1}.


We can note that a generic characteristic of the EU scenario is the
existence of a superinflation phase where $\dot{H} >0$ before
slow-roll inflation.
In this model the evolution described in Eq.~(\ref{a-EU1})
corresponds precisely to the superinflationary phase of the
evolution of the EU, which asymptotically approaches the de Sitter
expansion phase.


\begin{figure}
\centering
\includegraphics[width=8cm]{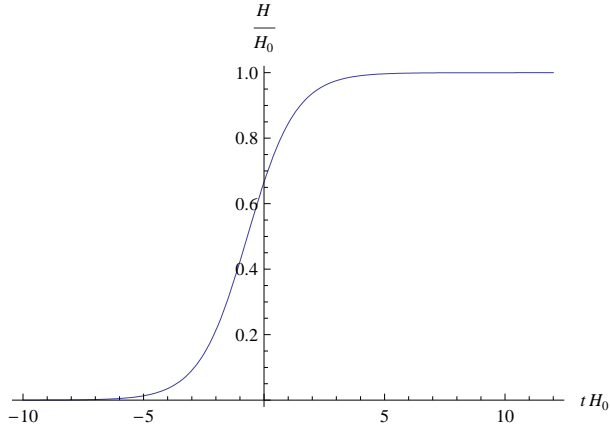}
\caption{Evolution of $H$ as a function of the cosmological time.
\label{H-dot}}
\end{figure}

Figure \ref{H-dot} shows a generic evolution of $H$ as a function of
the cosmological time obtained from Eq.~(\ref{a-EU1}).
We can note that $H$ increases with time from zero (the static
regimen) to the constant value $H_0$ (the onset of slow-roll
inflation).

The prediction of a superinflating phase before the onset of
slow-roll inflation arises in any EU scenario since the behavior of
$H$ as a function of the cosmological time depicted in
Fig.~\ref{H-dot} is general for all the EU models.

As was recently mentioned in the context of bouncing universes
\cite{Biswas:2013dry} and early in Ref.~\cite{Zhang:2003eh}, a
superinflation period could modify the spectral tilt at low $l$ by
making it blue-tilt before the conventional slow-roll inflation in
which a decreasing $H$ yields a red-tilted spectrum. This
possibility will be studied in the next section for the EU scenario.

The mechanism which generates this superinflationary period depends
on the particular EU model under consideration, but it is a generic
characteristic of the EU scenario. For example, in the models in
\cite{Ellis:2002we, Ellis:2003qz}, it is considered a closed
universe. Then, in this case, the spatial curvature is responsible
for the superinflationary period.

\section{The primordial perturbation in EU}\label{Perturbation}

The scalar perturbations to the FRW geometry, in the longitudinal
gauge, can be written as follows:

\begin{equation}
ds^2 =  (1+2\Phi)\,dt^2 - (1-2\Phi)\,a(t)^2\,d\vec{x}^2,
\end{equation}
where $\Phi$ is the Newtonian gravitational potential.

Normally in the emergent universe scenario, the Universe is
positively curved; see for example \cite{Ellis:2002we,
Ellis:2003qz}.
In this first approach to the problem, we have neglected the
contributions of the space curvature to the primordial perturbation,
but we have included a first approach to this point in Appendix A;
see also \cite{carlos}.

The equation for the perturbations in momentum space is given by

\begin{equation}\label{v1}
v_k'' +k^2 \,v_k - \frac{z''}{z}\,v_k =0 \,.
\end{equation}

where we have used the Mukhanov variable \cite{Mukhanov,
Garriga:1999vw},

\begin{equation}
v_k = a\left(\delta \phi_k +\frac{\phi'}{h}\,\Phi_k \right),
\end{equation}
where $\delta \phi_k$ are the perturbations in the inflaton field
and $'$ is derivative with respect to the conformal time $\eta =
\int dt/a$ and we have used units $M_p = (8 \pi G)^{-1/2}=1$. Also,
we have defined
\begin{eqnarray}
z &=& \frac{a\phi'}{h}\,,\\
h &=& \frac{a'}{a} \,.
\end{eqnarray}
%


In order to solve Eq.~(\ref{v1}), we consider the evolution of the
scale factor $a(t)$ given in Eq.~(\ref{a-EU1}) written in the
conformal time,
\begin{equation}\label{a-con}
a(\eta) = \frac{a_0}{1 -e^{a_0 H_0 \eta}}\,, \;\;\;\; \eta < 0\;\;.
\end{equation}

From Eq.~(\ref{v1})  and using Eq.~(\ref{a-con}), we obtain

\begin{equation}\label{pert-eq}
v''_k - (a_0\,H_0)^2 e^{a_0 H_0 \eta}\left(\frac{ \left(1+e^{a_0 H_0
\eta}\right)}{\left(-1+e^{a_0 H_0 \eta}\right)^2}\right)v_k + k^2v_k
=0\,,
\end{equation}
where we have considered $z''/z \approx a''/a$.
This equation is solved by

\begin{eqnarray}\label{sol1}
v_k(\eta) = \frac{1}{\sqrt{2k}}\left[\frac{e^{-ik\,\eta}}{1 - e^{a_0
H_0 \eta}}\right]&\times&\\
\nonumber\\
 _2F_1\Big(-1-\frac{i
k}{a_0H_0}-\sqrt{1-\left(\frac{ k}{a_0H_0}\right)^2},&-&1-\frac{i
k}{a_0H_0} + \sqrt{1-\left(\frac{ k}{a_0H_0}\right)^2}; 1-2\frac{i
k}{a_0H_0}; e^{a_0 H_0 \eta}\Big)\,,\nonumber
\end{eqnarray}
where $2F_1$ is the hypergeometric function.

In the solution (\ref{sol1}) we have considered (and appropriately
normalized) the solution of Eq. (\ref{pert-eq}) such that in the
short wavelength limit, the normalized positive frequency modes
correspond to the minimal quantum fluctuations,

\begin{equation}
v_k \approx \frac{e^{-ik\eta}}{\sqrt{2k}} \,,\;\;\; a\,H << k \;\;.
\end{equation}

Following \cite{Contaldi:2003zv}, we consider the spectrum of $Q
\equiv v/a$ which becomes constant at late time,

\begin{equation}\label{primordial}
P_Q = \frac{k^3}{2\pi^2}\big|Q\big|^2 \longrightarrow
\frac{H^2_0}{\pi^2}\,\frac{\chi^2\, \Gamma[x_1]\,
\Gamma[x_1^*]}{\Gamma[x_2]\, \Gamma[x_2^*]\, \Gamma[x_3]\,
\Gamma[x_3^*]}\,,
\end{equation}

where we have defined
\begin{eqnarray}
x_1 &=& 1- 2\,i \chi \\
x_2 &=& 2-i \chi-\sqrt{1-\chi^2}\\
x_3 &=& 2-i \chi + \sqrt{1-\chi^2} \\
\chi &=& \frac{k}{a_0\,H_0}
\end{eqnarray}

We can note that in the short wavelength limit($k >> H_0$), we
recovered the standard result of a nearly scale-invariant spectrum,

\begin{equation}
P_Q \rightarrow \left(\frac{H_0}{2\pi}\right)^2 \,.
\end{equation}

\begin{figure}
\centering
\includegraphics[width=10cm]{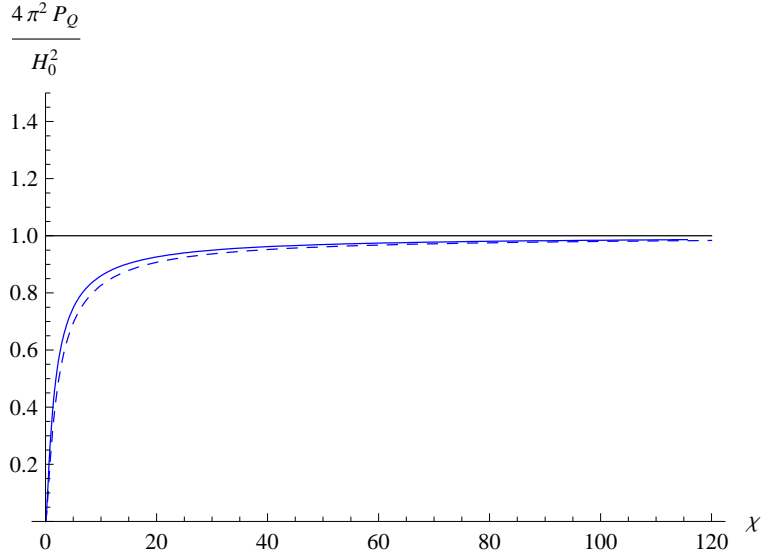}
\caption{Power spectrum for $Q$. The analytical computation
Eq.~(\ref{primordial}), solid line. The approximate spectrum, dashed
line Eq.~(\ref{aprox-P}). \label{P-Q}}
\end{figure}

At superhorizon scales, the two modes $Q_k$ and $\Phi_{k}$ are
related by a $k$-independent rescaling so that the spectrum given by
Eq.~(\ref{primordial}) directly translates into the spectrum of
$\Phi$. In Fig.~(\ref{P-Q}) we have plotted the spectrum of $P_Q$ as
a function of $\chi$ obtained from the analytical calculation
Eq.~(\ref{primordial}), solid line. We can note that there is a
suppression of the long wave modes as we expected given the
superinflationary regimen.


\section{The effective power spectrum}\label{Effective-Power}

In the last section we study the spectrum generated during the
superinflationary regimen. From this result and by following
Ref.~\cite{Liu:2013kea} we can estimate an effective power spectrum
generated by the EU model which takes into account the early period
of superinflation and the subsequent period of standard slow roll.
In this respect, from the result of previous section, we can note
that the scale invariance of the spectrum is the result of
inflationary evolution after the superinflationary stage. Then, if
the inflationary period is generated by the standard slow-roll
conditions, we can expect the usual slight red-tilted spectrum
during this period.
From Eq.~(\ref{primordial}) and by following
Ref.~\cite{Liu:2013kea}, we could model this situation by
considering the following scalar spectrum:

\begin{equation}\label{P_phi}
{\cal P}_\Phi = \,A\,\,k^{n_s -1}\!\frac{\chi^2\, \Gamma[x_1]\,
\Gamma[x_1^*]}{\Gamma[x_2]\, \Gamma[x_2^*]\, \Gamma[x_3]\,
\Gamma[x_3^*]}\,.
\end{equation}

For small $k$, this spectrum reproduces the behavior of spectrum
Eq.~(\ref{primordial}), i.e., suppression of the long wave modes. On
the other hand, when $k \gtrsim H_0\,a_0$, the spectrum reproduces
the usual slightly red spectrum of slow-roll inflation.

Another possibility is to consider a cutoff in the spectrum. This
cutoff is related to the transition from the superinflationary
regimen to slow-roll inflation which occurs when the mode $k_{max}$
exits the horizon, after which we would get the usual red-tilted
power law spectrum. In this case, we can write

\begin{equation}\label{P_cutoff}
{\cal P}_\Phi = \left\{
\begin{array}{l}
\bar{A}\, \frac{\chi^2\, \Gamma[x_1]\, \Gamma[x_1^*]}{\Gamma[x_2]\,
\Gamma[x_2^*]\, \Gamma[x_3]\, \Gamma[x_3^*]}\,\,,\;\;\; k<k_{max} \\
\\
A\,\left(\frac{k}{k_{max}}\right)^{n_s -1}\,,\;\;\; k>k_{max}
\end{array} \right.
\end{equation}

The constants $A$ and $\bar{A}$ are chosen in order to match the
superinflationary power spectrum Eq.~(\ref{primordial}) with the
slow-roll inflationary power spectrum at $k= k_{max}$.


\section{CMB Anisotropies}\label{CMB-Anisotropies}

We compute the theoretical CMB power spectrum of this model by using
CLASS code \cite{Lesgourgues:2011re, Blas:2011rf}.
In this first approximation, in the context of the EU scenario, for
the study of the consequences of the superinflationary phase  to the
CMB power spectrum, we have simplified the code for computing the
$C_l$ by considering an approximation of the spectrums
Eqs.~(\ref{P_phi}) and (\ref{P_cutoff}).
In particular, we approximate the spectrum of the superinflation
phase Eq.~(\ref{primordial}) with the following expression (see for
example \cite{Biswas:2013dry}),

\begin{equation}\label{aprox-P}
P_Q \sim \frac{H_0^2}{4\pi^2}\,\frac{\chi^2}{\left(1 + \chi
\right)^2}\,,
\end{equation}
which is obtained by considering the evolution of the scale factor
given by Eq.(\ref{a-EU1}) and the Hubble crossing condition,

\begin{equation}
k = a\,H \;\;\Rightarrow \; k = H_0\,A\,e^{H_0t} \,.
\end{equation}

\begin{figure}
\centering
\includegraphics[width=8cm, angle=-90]{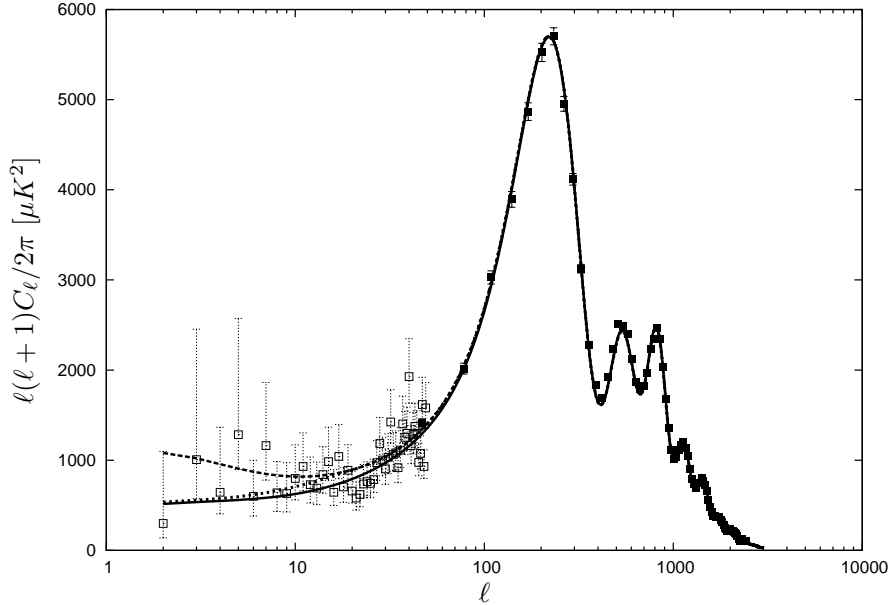}
\caption{Temperature power spectrum for the pure power law (dashed
line), EU scenario (solid line), and EU scenario with a cutoff
(points line). The points show the Planck data. \label{Cls}}
\end{figure}

In Fig. \ref{P-Q}  we have plotted the spectrum Eq.~(\ref{aprox-P}),
where we can compare it with the analytic calculation
Eq.~(\ref{primordial}). We can note that the approximated spectrum
reproduces very well the  spectrum obtained by the analytic
calculation.


In order to show the suppression at large scales coming from the EU
scenario, we have plotted in Fig. \ref{Cls} the temperature power
spectrum obtained from the pure power law (dashed line), from the EU
scenario (solid line), and from the EU scenario with a cutoff
(dotted line). The points show the Planck data. We can note that
compared to the standard power-law model, the $C_l$ spectrum in the
emergent universe scenario is suppressed at large scales.

In these examples, we have considered the following values for the
parameters in the case of, the EU scenario in Eqs.~(\ref{P_phi}) and
(\ref{aprox-P}):  $A= 2.07 \times 10^{-9}$, $n_s=0.9603$, and
$a_0\,H_0 = 0.0002\, \textrm{Mpc}^{-1}$. In the case of the EU
scenario with the cutoff Eqs.~(\ref{P_cutoff}) and (\ref{aprox-P}),
we consider $A= 2.42 \times 10^{-9}$, $n_s=0.967$, $a_0\,H_0 =
0.0003\, \textrm{Mpc}^{-1}$, and $k_{max}=0.0015\,
\textrm{Mpc}^{-1}$. At this moment we are not doing a best-fit
calculation of theses parameters, just showing two particular cases
and how they produce a suppression of the spectrum at large scales.

We have to mention that there is a tuning between the turn-around
from superinflation to slow-roll inflation and when the comoving
scale corresponding to the current horizon crosses the Hubble radius
during inflation. This is typical for trying to account for the
suppression of the low multipoles due to preinflationary dynamics;
see for example \cite{Contaldi:2003zv,Schwarz:2009sj,
Ramirez:2011kk, Dudas:2012vv, Ramirez:2012gt, Cicoli:2014bja}.
In particular, in this work we consider that this scale, which is
determined by the combination ($a_0\,H_0$), is a free parameter.
However, there is the possibility of linking this scale with the
stability conditions of the ES solution. In this case, we could
obtain bounds on this scale which could be related with the current
horizon scale, similar to the case where stability conditions of the
ES solution impose an upper bound on the cosmological constant
\cite{Guendelman:2013dka}. We expect to return to this point in
future work.

\section{Conclusions}\label{Sec-conclusions}

In recent cosmological observations, there are intriguing results on
the cosmic microwave background radiation suggesting a lack of power
at large angular scales. This situation is interesting because it
may give us clues towards physics at the very early Universe,
perhaps before slow-roll inflation.
In this context, the EU models are an interesting arena to explore
preinflationary physics and its possible implications for the CMB
anomalies.

In this paper we study the primordial perturbations in the context
of the EU scenario. In particular, we focus on the primordial
perturbations generated during the superinflationary phase.
We find that the superinflationary period in the EU scenario
produces a suppression of the primordial perturbations at large
scale which could be responsible for the observed lack of power at
large angular scales of the CMB.

In particular, in this work we considered the EU model developed in
Ref.~\cite{Ellis:2002we, Ellis:2003qz} as an arena in which to
explore the consequences of the superinflationary regimen.
We calculated the primordial perturbations generated during the
superinflationary phase and the effective power spectrum generated
by the EU model, which take into account the early period of
superinflation and the subsequent period of standard slow-roll
inflation.
By using the CLASS code, we compute the theoretical CMB power
spectrum generated by this model and compare it with the standard
results of slow-roll inflation and the Planck data.

In this first approach to the problem, we have neglected the
contributions of the space curvature to the primordial perturbation
(most of the EU models consider closed universes), but we have
included a first approach to this point in Appendix A.

In this model, the scale of transition from superinflation to
slow-roll inflations is determined by the combination ($a_0\,H_0$),
which is considered a free parameter.
Preliminary results show that the global behavior (suppression of
the CMB anisotropies at large scale and agreement with Planck data
at large $l$) is not particulary sensitive to the election of this
combination once the scale $a_0\,H_0$ is smaller than a typical
scale of order $\approx 0.002\, \textrm{Mpc}^{-1}$.
However, there is the possibility of linking this scale with the
stability conditions of the ES solution.
In this first approach to the problem, we have not considered a
best-fit calculation of the parameters of the model. We expect to
return to these points in the near future by including a best-fit
calculation and by considering bounds on the free parameters of the
model that come from stability conditions of the ES solution; see
\cite{Guendelman:2013dka}.

In summary, in this work we show that the superinflation phase,
which is a characteristic shared by all EU models, could be
responsible for part of the anomaly in the low multipoles of the
CMB, in particular, for the observed lack of power at large angular
scales.


\section{acknowledgments}

P. L. has been partially supported by FONDECYT grant N$^{0}$
11090410, Mecesup UBB0704, and Universidad del B\'{i}o-B\'{i}o
through Grants GI121407/VBC and 141407 3/R. We are grateful to I.
Duran for collaboration with the CLASS code. We thank J. Garriga and
D. Espriu for useful discussion and  A. Cid for providing comments
on the manuscript.

\section*{Appendix A}\label{app1}


In this appendix we study the period of superinflation in the
specific EU model developed in Ref.~\cite{Ellis:2002we,
Ellis:2003qz}. This EU model is based on general relativity and
considered a closed universe dominated by a scalar field minimally
coupled to gravity.

The energy density, $\rho$, and the pressure, $P$, are expressed by
the following equations,
\begin{eqnarray}
\rho &=& \frac{1}{2}\dot{\phi}^2 + V(\phi) \,,\\
p &=& \frac{1}{2}\dot{\phi}^2 - V(\phi) \,,
\end{eqnarray}
where $V(\phi)$ is the scalar potential show in
Fig.\ref{Potencial-1}.
The Friedmann and the Raychaudhuri field equations become

\begin{eqnarray}
H^2 = \frac{8\pi G}{3}\left(\frac{1}{2}\dot{\phi}^2 + V(\phi)
\right) - \frac{1}{a^2}, \label{H} \\
\nonumber\\
\ddot{\phi} +3H\,\dot{\phi} = -\frac{\partial
V(\phi)}{\partial\phi}\,. \label{phi}
\end{eqnarray}

The other nonindependent equation is
\begin{eqnarray}\label{EC-a}
\dot{H} = - 4\pi G\, \dot{\phi}^2 + \frac{1}{a^2}\label{H2}.
\end{eqnarray}

During the past-eternal static regime, it is assumed that the scalar
field is rolling on the asymptotically flat part of the scalar
potential with a constant velocity, providing the conditions for a
static universe. But once the scalar field exceeds some value, the
scalar potential slowly droops from its original value. The overall
effect of this is to distort the equilibrium behavior breaking the
static solution. If the potential has a suitable form in this
region, superinflation and slow-roll inflation will occur.
In this case, the evolution of the scale factor could be modeled by
Eq.~(\ref{a-EU1}); see Ref.~\cite{Ellis:2003qz} and the discussion
in Sec. \ref{EU}.
%
%

From Eq.~(\ref{EC-a}) we can note that, in this model, the mechanism
which generated the superinflationary period is the effect of the
curvature of the closed universe.


The scalar perturbations to the closed FRW geometry, in the
longitudinal gauge, can be written as follows,

\begin{equation}
ds^2 =  (1+2\Phi)\,dt^2 - (1-2\Phi)\,a(t)^2\left[\frac{dr^2}{1-r^2}
+ r^2\,(d\theta^2 + \sin^2 \!\!\theta\,\, d\phi^2 )\right],
\end{equation}
where $\Phi$ is the Newtonian gravitational potential.


The equation for the perturbations in momentum space is similar to
the one discussed in Sec.~\ref{Perturbation} and it is given by

\begin{equation}\label{v1_a}
v_k'' +k^2 \,v_k - \frac{z''}{z}\,v_k =0 \,.
\end{equation}
where we have considered that the scalar potential is nearly
constant, and we have neglected terms proportional to $dV/d\phi$.
Also, we have used the Mukhanov variable \cite{Mukhanov,
Garriga:1999vw}

\begin{equation}
v_k = a\left(\delta \phi_k +\frac{\phi'}{h}\,\Phi_k \right),
\end{equation}
where $\delta \phi_k$ is the perturbation in the inflaton field and
$'$ is derivative with respect to the conformal time $\eta = \int
dt/a$, and we have used units $M_p = (8 \pi G)^{-1/2}=1$. Also, we
have defined
\begin{eqnarray}
z &=& \frac{a\phi'}{h}\left(1 +\frac{3}{\Delta}\right)^{1/2}\,,\\
h &=& \frac{a'}{a} \,.
\end{eqnarray}

Here, the Laplacian should be understood as a c number, representing
the corresponding eigenvalue \cite{Garriga:1999vw}.


In order to solve Eq.~(\ref{v1_a}), we consider the evolution of the
scale factor $a(t)$ given in Eq.~(\ref{a-EU1}) written in the
conformal time Eq.~(\ref{a-con}). Then, from Eq.~(\ref{v1_a}) and by
using Eq.~(\ref{a-con}), we obtain the following equation,

\begin{equation}\label{pert-eq-a}
v''_k - (a_0\,H_0)^2 e^{a_0 H_0 \eta}\left(\frac{ \left(1+e^{a_0 H_0
\eta}\right)}{\left(-1+e^{a_0 H_0 \eta}\right)^2}\right)v_k + k^2v_k
=0\,,
\end{equation}
where we have considered the eigenvalues of the Laplacian operator
for a closed space (see \cite{Abbott:1986ct}):
\begin{eqnarray}\label{delta_cerrado1-a}
\Delta v_k &=& -k^2 \,v_k \\
&=& - (\beta^2 -1)\,v_k \,.\label{delta_cerrado2-a}
\end{eqnarray}

In this case $\beta = 3,4, 5, \dots$. The modes $\beta = 1, 2$ are
pure gauge modes and are not included in the spectrum
\cite{Abbott:1986ct}. Also, we have considered $z''/z \approx
a''/a$.

Equations (\ref{pert-eq-a}) are solved by Eq.~(\ref{sol1}), where
$k$ is now defined in Ecs.~(\ref{delta_cerrado1-a}) and
(\ref{delta_cerrado2-a}).
From this solution and by following a similar procedure as in
Sec.~\ref{Perturbation}, we found the spectrum of primordial
perturbations. We can note that the qualitative behavior of these
primordial perturbations is similar to that discussed in
Sec.~\ref{Perturbation}: a flat spectrum for short wavelengths and a
suppression of the long wave modes, as we expect given the
superinflation regimen.

\end{document}